\documentclass[11pt,a4paper]{article}

\usepackage{amsmath}
\usepackage{mathtools}
\usepackage{nccmath}
\usepackage{amsthm}
\usepackage{amssymb}
\usepackage{amsfonts}
\usepackage{anyfontsize}
\usepackage{mathrsfs}
\usepackage{bbm}
\usepackage{graphicx}
\usepackage{caption}
\usepackage{bookmark,hyperref}
\usepackage{12many}
\usepackage{parskip}
\usepackage{setspace}
\usepackage{multirow}
\usepackage{arydshln}
\usepackage{anyfontsize}
\usepackage{changepage}
\usepackage{todonotes}
\usepackage{braket}
\usepackage{dirtytalk}
\usepackage{MnSymbol}

\newcommand{\psymbol}[2]{\genfrac{}{}{0pt}{1}{#1}{#2}}

\numberwithin{equation}{section}
\usepackage{fullpage}

\begin{document}

\begin{titlepage}
 \thispagestyle{empty}
 \begin{flushright}
 \end{flushright}

 \vspace{30pt}

 \begin{center}
     
  {\fontsize{20}{24} \bf {Asymmetric $\mathbb{Z}_4$ orbifolds of type IIB \\ \vspace{0.7cm} string theory revisited
  \vspace{20pt} }}

     \vspace{20pt}
{\fontsize{13}{16}\selectfont {George Gkountoumis} \\[10mm]}

{\small\it
 Institute for Theoretical Physics {and} Center for Extreme Matter and Emergent Phenomena \\
Utrecht University, 3508 TD Utrecht, The Netherlands }

\vspace{1cm}

{\bf Abstract}

\vspace{0.3cm}

\begin{adjustwidth}{12pt}{12pt}

We construct freely acting asymmetric $\mathbb{Z}_4$ orbifolds of type IIB string theory on $T^5$ preserving 24,16 or 8 supercharges in five dimensions. We show that these models are well-defined if the SO(8) lattice is chosen, instead of the SU(2)$^4$ lattice, which was previously considered in the literature.

\end{adjustwidth}

\end{center}

\vspace{5pt}

\newcommand\blfootnote[1]{%
  \begingroup
  \renewcommand\thefootnote{}\footnote{#1}%
  \addtocounter{footnote}{-1}%
  \endgroup}

\blfootnote{g.gkountoumis@uu.nl}

\noindent

\end{titlepage}

\begin{spacing}{1.15}
\tableofcontents
\end{spacing}

\section{Introduction}

Freely acting asymmetric orbifolds of type IIB string theory preserving 24,16 or 8 supercharges in five dimensions have been recently discussed in \cite{Gkountoumis:2023fym} (for some earlier references  see e.g. \cite{Rohm:1983aq,Kounnas:1988ye,Ferrara:1987es,Ferrara:1988jx,Kiritsis:1997ca}). The orbifolds discussed in this paper have target spaces of the form $\mathbb{R}^{1,4}\times T^5$ identified under the action of a $\mathbb{Z}_p$ symmetry. Our interest lies in the same class of orbifolds and, in particular, on $\mathbb{Z}_4$ models.

Our motivation for studying $\mathbb{Z}_4$ asymmetric orbifolds comes from recent work on models preserving 24 supercharges \cite{bianchi2022perturbative,Gkountoumis:2023fym} (for some other references see e.g. \cite{Ferrara:1989nm,Bianchi:2008cj,Bianchi:2010aw}). In the class of orbifolds mentioned above, there are in fact only four candidate models preserving 24 supercharges in five dimensions, that is $\mathcal{N}=6$ supersymmetry in $D=5$, and these are orbifolds of rank 2,3,4 and 6. Orbifolds of rank 2 and 3
were constructed in \cite{bianchi2022perturbative,Gkountoumis:2023fym}, where it was also argued that the rank 4 and 6 orbifolds cannot be consistently constructed, as they lead to non-integer degeneracy of bosonic ground state in the twisted sectors, which is not allowed \cite{Narain:1986qm,narain1991asymmetric}.

However, in this note we show that the $\mathbb{Z}_4$, $\mathcal{N}=6$  model can be consistently constructed and that the appropriate lattice 
to consider is the $D_4$ lattice, instead of the $(A_1)^4$ lattice that was previously
discussed in the literature\footnote{$D_n$ and $A_n$ denote the root lattices of $\text{SO}(2n)$ and $\text{SU}(n+1)$ respectively. $(A_1)^4$ is a shorthand notation for $A_1 \oplus A_1 \oplus A_1 \oplus A_1$.}. We construct explicitly the modular invariant partition function for the model based on the $D_4$ lattice, and we analyze the spectrum of lowest excited string states in the untwisted and twisted orbifold sectors.

Finally, by applying slight modifications to the $\mathbb{Z}_4$, $\mathcal{N}=6$ orbifold, we construct models preserving 16 or 8 supercharges in five dimensions. Regarding the rank 6 model, we did not find a consistent construction.

\section{Freely acting asymmetric orbifolds}
\label{section orbifolds}
In this section we first discuss briefly freely acting asymmetric orbifolds of type IIB string theory on $T^5$; the discussion is mostly based on \cite{Gkountoumis:2023fym, Gkountoumis:2024dwc}. Then we present three $\mathbb{Z}_4$ asymmetric freely acting orbifolds preserving 24,16 or 8 supercharges in five dimensions. We mostly focus on the model preserving 24 supercharges, as for the construction of models preserving less supersymmetry only slight modifications are required.

\subsection{General remarks and consistency conditions}
\label{general remarks}
For the construction of asymmetric orbifolds we follow the procedure presented in the original papers \cite{Narain:1986qm,narain1991asymmetric}. In general, upon compactification on $T^5$ the momentum and winding numbers take values in a Narain lattice $\Gamma^{5,5}$, which is an even, self-dual lattice \cite{narain1989new}. For our purposes we decompose $T^5=T^4\times S^1$ and correspondingly decompose the lattice as $\Gamma^{4,4}(\mathcal{G})\oplus \Gamma^{1,1}$, where $\Gamma^{4,4}(\mathcal{G})$ is an even, self-dual Lie algebra lattice admitting purely left and right-moving  symmetries. Such a lattice can be realised at special points in the moduli space as 
\begin{equation}
    \Gamma^{4,4}(\mathcal{G}) \equiv \{(p_L,p_R)|\,p_L \in \Lambda_W(\mathcal{G}),\, p_R \in \Lambda_W(\mathcal{G}),\, p_L-p_R \in \Lambda_R(\mathcal{G})\}\,.
\end{equation}
Here $\mathcal{G}$ is a Lie algebra of rank four and $\Lambda_W(\mathcal{G})$, $\Lambda_R(\mathcal{G})$ denote the weight and root lattices of $\mathcal{G}$, respectively. Now, the orbifold acts as a rotation on $\Gamma^{4,4}(\mathcal{G})$ and as a shift on $\Gamma^{1,1}$. Here we will only consider rotations $\mathcal{M}_{\theta}=(\mathcal{N}_L,\mathcal{N}_R) \in \text{SO}(4)_L\times \text{SO}(4)_R \subset \text{SO}(4,4)\,$ that do not mix left and right-movers. 
For a $\mathbb{Z}_p$ orbifold, we require that the rotation satisfies $(\mathcal{M}_{\theta})^p=1$. Consistency of the asymmetric orbifold requires that the rotation is in the automorphism group of the lattice\footnote{A discussion on consistent rotations can be found e.g.\ in \cite{lerchie1989lattices}, cf. appendix B.}. Since the orbifold acts as a shift on $\Gamma^{1,1}$, it leaves $\Gamma^{1,1}$ invariant\footnote{Due to the shift, momentum states pick up a phase in the untwisted sector. In the twisted sectors states become massive.}. On the other hand, $\Gamma^{4,4}(\mathcal{G})$ is not in general invariant under rotations $\mathcal{M}_{\theta}$. If there exists a sublattice $I\subset \Gamma^{4,4}(\mathcal{G})$ that is invariant under the orbifold action, it is given by
\begin{equation}
    I \equiv \{ p \in \Gamma^{4,4}(\mathcal{G})\,|\, \mathcal{M}_{\theta}\cdot p = p\}\,.
\end{equation}
Then the complete  sublattice that is invariant under the 
orbifold action is 
\begin{equation}
    \hat{I}=I\oplus \Gamma^{1,1}\,.
\end{equation}
It will be useful to determine the orbifold action on the world-sheet fields. We denote the $S^1$ coordinate by $Z$ and the  four real $T^4$ coordinates by $Y^m, m=1,\ldots 4$, which we combine into two complex coordinates
 $W^i = \tfrac{1}{\sqrt{2}}(Y^{2i-1}+iY^{2i})$ with $i=1,2$. We parametrize the rotations $(\mathcal{N}_L,\mathcal{N}_R) \in \text{SO}(4)_L\times \text{SO}(4)_R \subset \text{SO}(4,4)$ by two twist vectors $\tilde{u}=(0,0,\tilde{u}_3,\tilde{u}_4)$ and $u=(0,0,u_3,u_4)$ as follows
\begin{equation}
    \mathcal{N}_L = \begin{pmatrix}
        R(2\pi\tilde{u}_3)&0\\
        0&R(2\pi\tilde{u}_4)
        \end{pmatrix}\,, \qquad \mathcal{N}_R = \begin{pmatrix}
        R(2\pi u_3)&0\\
        0&R(2\pi u_4)
        \end{pmatrix}\,,
        \label{orbifold twist in terms of twist vectors}
\end{equation}
where we use the notation $R(x)=\begin{psmallmatrix}\cos x & \,\,\,\,-\sin x \\ \sin x & \,\,\,\,\cos x \end{psmallmatrix}$ for a $2\times 2$ rotation matrix\footnote{In order to properly define the orbifold action on all the fields, $\mathcal{N}_{L/R}$ should be uplifted to Spin$(4)_{L/R}$; for more details see \cite{Gkountoumis:2023fym}.}. Then, the orbifold acts on the bosonic torus coordinates through (asymmetric) rotations
\begin{equation}\label{orbiaction2}
\begin{aligned}
{W}_{L}^1 \;&\rightarrow\; e^{2\pi i\tilde{u}_3}\: {W}_{L}^1 \,, \\
{W}_{L}^2 \;&\rightarrow\; e^{2\pi i\tilde{u}_4}\: {W}_{L}^2 \,, \\
W_{R}^1 \;&\rightarrow\; e^{2\pi i{u}_3}\: W_{R}^1 \,, \\
W_{R}^2 \;&\rightarrow\; e^{2\pi i{u}_4}\: W_{R}^2 \,,
\end{aligned}
\end{equation}
and with the same action on their world-sheet superpartners. The rotations on the torus are accompanied by a shift along the circle coordinate
\begin{equation}\label{shift}
Z \;\rightarrow\; Z + 2\pi \mathcal{R} / p \,,
\end{equation}
which makes the orbifold freely acting. Here $\mathcal{R}$ denotes the radius of the circle.

Of course, we have to ensure that our models satisfy modular invariance. This can be verified if the following conditions hold \cite{vafa1986modular,Baykara:2023plc}
\begin{equation}
    p \sum_{i=3}^4 \tilde{u}_i \,\in\, 2\mathbb{Z} \qquad\text{and}\qquad p \sum_{i=3}^4 u_i \,\in\, 2\mathbb{Z} \,,
    \label{modular conditions on twist vectors}
\end{equation}
where $p$ is the orbifold rank. Also, if the rank of the orbifold is even, we check the additional condition\footnote{In some cases it is possible to construct consistent orbifolds even if this condition is not satisfied \cite{Harvey:2017rko}.} for $(p_L,p_R)\in \Gamma^{4,4}(\mathcal{G})$ \begin{equation}\label{modcond2}
   p_L \mathcal{N}_L^{\,p/2} p_L - p_R \mathcal{N}_R^{\,p/2} p_R \,\in\, 2\mathbb{Z}\,. 
\end{equation}

\subsection{$\mathcal{N}=6$ model}
\label{the rank 4 model}
In this section we consider a $\mathbb{Z}_4$ freely acting asymmetric orbifold of type IIB string theory on $S^1\times T^4$ preserving 24 supercharges in five dimensions, that is $\mathcal{N}=6$ in $D=5$. The twist vectors for this orbifold are given by
\begin{equation}
    \tilde{u}=\left(0,0,\frac{1}{4},\frac{1}{4}\right)\ ,\qquad u=\left(0,0,0,0\right)\,.
    \label{twist vectors Z4}
\end{equation}
It can be easily verified that the above twist vectors satisfy \eqref{modular conditions on twist vectors} (for this orbifold $p=4$). The appropriate lattice to begin with is
\begin{equation}
    \Gamma^{5,5}=\Gamma^{4,4}(D_4)\oplus\Gamma^{1,1}\,.
\end{equation}
Now, we check the condition \eqref{modcond2}. We have
\begin{equation}
    p_L\mathcal{N}_L^{\,2}p_L - p_R\mathcal{N}_R^{\,2}p_R=-\left(p_L^2+p_R^2\right)\, ,
\end{equation}
which is even for $p_L-p_R \in \Lambda_R(D_4)$. So, modular invariance is ensured. 

The invariant sublattice $I$ is spanned by the vectors $(0,p_R)$, which combined with the condition ${p}_L-{p}_R\in \Lambda_R(D_4)$ yields ${p}_R\in \Lambda_R(D_4)\subset \Lambda_W (D_4)$. Recall that the orbifold acts as a shift on $\Gamma^{1,1}$ and leaves this lattice invariant. So, the complete orbifold invariant lattice is
\begin{equation}
    \hat{I}=\Lambda_R(D_4) \oplus\Gamma^{1,1}\ .
\end{equation}
Let us now move on to the spectrum of our model. We focus on the lowest excited string states; these are either massless states or the lightest massive states. The spectrum can be read off from the partition function, and all the details can be found in appendix \ref{spectrumapp}. We start from the untwisted sector. The massless spectrum consists of the graviton, 15 vectors, 14 scalars, 6 gravitini and 20 dilatini. These massless fields fit into the $\mathcal{N}=6$ gravity multiplet in $D=5$ \cite{Cremmer:1980gs}. The 14 massless scalars parameterize the moduli space
\begin{equation}
    {\cal M}_{\mathcal{N}=6}=\frac{\text{SU}^*(6)}{\text{USp}(6)}\ .
\end{equation}
Now, we present the spectrum of the lightest massive states; for the massive states we also write down their representations under the massive little group $\text{SU}(2)\times \text{SU}(2)$ in five dimensions. We find 2 gravitini (\textbf{3},\textbf{2}), 4 tensors (\textbf{3},\textbf{1}), 8 vectors (\textbf{2},\textbf{2}), 26 dilatini, $16\times(\textbf{2},\textbf{1})$ and $10\times(\textbf{1},\textbf{2})$, and 20 scalars (\textbf{1},\textbf{1}). All these fields have mass $\left|1/\mathcal{R}\right|$,\footnote{In a non-freely acting orbifold these states would have been projected out of the spectrum. Here, in the freely acting orbifold, these states survive the orbifold projection with the addition of momentum modes along the circle. This makes the states massive.} and fit into a complex (1,2) BPS supermultiplet in the representations
\begin{equation}
(\textbf{3},\textbf{2})\oplus2\times(\textbf{3},\textbf{1})\oplus4\times(\textbf{2},\textbf{2})\oplus5\times(\textbf{1},\textbf{2})\oplus8\times(\textbf{2},\textbf{1})\oplus10\times(\textbf{1},\textbf{1})\,.
\label{bps reps n=6}
\end{equation}
We can also construct Kaluza-Klein towers along the $S^1$ by adding appropriate momentum modes to all states. Then, the orbifold untwisted spectrum matches exactly with the Scherk-Schwarz  supergravity one (\cite{Scherk:1978ta,Scherk:1979zr,Cremmer:1979uq}) found in \cite{Hull:2020byc}.

We continue with the spectrum in the twisted sectors, which we label by $k$\footnote{The untwisted sector is denoted by $k=0$, and the twisted sectors by $k=1,\ldots,p-1$, where $p$ is the orbifold rank.}. Here, for clarity, we will write the spectrum that we find in the NS-NS, NS-R, R-NS and R-R sectors separately. 

We start from the $k=1$ sector.  In the NS-NS sector we find 2 vectors  (\textbf{2},\textbf{2}) and 8 scalars  (\textbf{1},\textbf{1}). In the NS-R sector we find 8 dilatini, $4\times(\textbf{2},\textbf{1})$ and $4\times(\textbf{1},\textbf{2})$. In the R-NS sector we find 1 gravitino (\textbf{3},\textbf{2}) and 5 dilatini, $4\times(\textbf{2},\textbf{1})$ and $1\times(\textbf{1},\textbf{2})$. Finally, in the R-R sector we find 2 tensors (\textbf{3},\textbf{1}), 2 vectors (\textbf{2},\textbf{2}) and 2 scalars  (\textbf{1},\textbf{1}).  All these fields have mass $\left|{\mathcal{R}}/{4\alpha'}\right|$, due to the fractional $\tfrac{1}{4}$-winding on the circle\footnote{In the twisted sectors of freely acting orbifolds the winding numbers on the circle are shifted by $k/p$.}. The spectrum in the $k=1$ sector coincides with the one from the $k=3$ sector. In total, from both $k=1$ and $k=3$ sectors we find 2 gravitini (\textbf{3},\textbf{2}), 4 tensors (\textbf{3},\textbf{1}), 8 vectors (\textbf{2},\textbf{2}), 26 dilatini, $16\times(\textbf{2},\textbf{1})$ and $10\times(\textbf{1},\textbf{2})$, and 20 scalars (\textbf{1},\textbf{1}). All these fields have mass $\left|{\mathcal{R}}/{4\alpha'}\right|$, and fit into a complex (1,2) BPS supermultiplet, as in \eqref{bps reps n=6}.

In the $k=2$ sector, the lightest  states that survive the orbifold projection carry a suitable combination of momentum modes, winding modes, and $T^4$ right-moving momenta\footnote{I would like to thank Guillaume Bossard for useful correspondence on this point.}. These are higher excited stringy states, which we omit writing down. This concludes our discussion on the spectrum of the $\mathbb{Z}_4$, $\mathcal{N}=6$ orbifold in $D=5$. Note that further compactification on $S^1$ yields a consistent model preserving 24 supercharges in four dimensions.

We conclude this section by comparing the spectrum of the $\mathbb{Z}_4$ freely acting orbifold with the non-freely acting orbifold $T^4/\mathbb{Z}_4 \times S^1$. In this case, there are no massive fields in the untwisted sector; these are projected out of the spectrum instead. In addition to the 6 massless gravitini from the untwisted sector, there are another 2 massless gravitini coming from the $k=1$ and $k=3$ twisted sectors, as now these states do not carry fractional winding numbers. This leads to a supersymmetry enhancement from 24 to 32 supercharges. In the $k=2$ sector all states are massive even without the fractional windings. Hence, we see that we simply obtain an $\mathcal{N}=8$ theory in five dimensions.

\subsection{$\mathcal{N}=4$ and $\mathcal{N}=2$ models}
By applying slight modifications to the $\mathcal{N}=6$, $\mathbb{Z}_4$ orbifold based on the lattice $\Gamma^{4,4}(D_4)\oplus\Gamma^{1,1}$, we can construct models preserving less supersymmetry in five dimensions. The strategy will be to choose the same lattice and slightly change the twist vectors.

For example, consider the $\mathbb{Z}_4$ orbifold with twist vectors $\tilde{u}=\left(0,0,\tfrac{1}{4},\tfrac{3}{4}\right)$ and $u=(0,0,0,0)$. This orbifold preserves 16 supercharges in five dimensions, and its spectrum consists of the $\mathcal{N}=4$, $D=5$ gravity multiplet and one vector multiplet\footnote{In order to find the spectrum of the models presented in this section we use Table 9 of \cite{Gkountoumis:2023fym}.}. The moduli space is fixed by supersymmetry and  is given by \cite{Awada:1985ep}
\begin{equation}
{\cal M}_{\mathcal{N}=4}=\mathbb{R}^+\times \frac{\text{SO}(5,1)}{\text{SO}(5)}\ ,
\end{equation}
where $\mathbb{R}^+=\text{SO}(1,1)/\mathbb{Z}_2$. Note that the number of vector multiplets is odd. This is in agreement with the results in \cite{Gkountoumis:2023fym}, where it was observed that in a very broad class of Minkowski string vacua preserving 16 supercharges in $5D$,\footnote{Higher dimensional models preserving 16 supercharges are classified in e.g. \cite{de2001triples,bedroya2022compactness,fraiman2023unifying}. All these models, upon dimensional reduction, yield an odd number of vector multiplets in $5D$.} only an odd number of vector multiplets arises.

As another example, we could combine the the twist vector $\tilde{u}=\left(0,0,\tfrac{1}{4},\tfrac{1}{4}\right)$ with a right-moving spacetime fermionic twist, that is $u=(0,0,0,1)=(-1)^{F_R}$. This is an  orbifold preserving 8 supercharges in five dimensions, and its spectrum consists of the $\mathcal{N}=2, D=5$ gravity multiplet coupled to six vector multiplets and no hypermultiplets. This orbifold falls into a class of models that were recently discussed in \cite{Baykara:2023plc,Gkountoumis:2024dwc}, as it has no hypermultiplets, and all massless fields come from the NS-NS sector\footnote{Earlier work on models preserving 8 supercharges can be found in e.g. \cite{Dolivet:2007sz,Anastasopoulos:2009kj,Condeescu:2013yma,Hull:2017llx,Gautier:2019qiq,Israel:2013wwa}.}. The appearance of only NS-NS scalars simplifies the computation of the classical moduli space. Using the results of \cite{Gkountoumis:2024dwc}, we find that for the model at hand, the 6 massless scalars parameterize the classical moduli space\footnote{This space can be also obtained from a truncation of the $\mathcal{N}=6$ theory \cite{Gunaydin:1983rk}.}
\begin{equation}\label{modspace1}
    {\cal M}_{\mathcal{N}=2}=\mathbb{R}^+\times \frac{\text{SO}(5,1)}{\text{SO}(5)}\ .
\end{equation}
Here the $\mathbb{R}^+$ factor is parametrized by the five-dimensional string coupling. Hence, the above moduli space could receive quantum corrections (see e.g. \cite{Antoniadis:2003sw,Robles-Llana:2006vct,Robles-Llana:2006hby,Robles-Llana:2007bbv,Alexandrov:2008gh,Alexandrov:2021shf,Alexandrov:2021dyl,Alexandrov:2023hiv}). However, as we have already mentioned, this orbifold fits in the class of models discussed in \cite{Gkountoumis:2024dwc}, where it was argued that for this class of models\footnote{These are freely acting asymmetric orbifold of type IIB string theory on $T^5$ preserving 8 supercharges in $5D$, in which all moduli come from the NS-NS sector.} the moduli spaces receive no quantum corrections.

Finally, we mention that compactification of the above models on $S^1$ leads to consistent models preserving $\mathcal{N}=4$ or $\mathcal{N}=2$ supersymmetry in $D=4$.

\section{Discussion}
In this note we first constructed a rank 4 freely acting asymmetric orbifold of type IIB string theory on $T^5$, based on the $D_4$ lattice, which preserves 24 supercharges in five dimensions. Then, by slightly modifying this model we also constructed models, based on the same lattice, preserving 16 or 8 supercharges in five dimensions.

\section*{Acknowledgements}
I am grateful to my advisor, Stefan Vandoren, for encouraging me to initiate this work, and for helpful discussions and comments on the draft. I would also like to thank  Massimo Bianchi, Guillaume Bossard and  Chris Hull for useful correspondence and comments on the draft.

\appendix
\label{appendix}

\section{Details of the $\mathcal{N}=6$ model}

\subsection{Partition function}
\label{Partition function with $D4$ lattice}
Here we present the partition function of the $\mathbb{Z}_4$ model discussed in section \ref{the rank 4 model}. For the construction of the partition function we follow \cite{Gkountoumis:2023fym}. Our conventions and various identities relevant for the construction of the partition function can be found in appendix \ref{modular functions}.

The orbifold partition function takes the form
\begin{equation}
    Z(\tau,\bar \tau)=\frac{1}{4}\sum_{k,l=0}^{3}{Z}[k,l](\tau,\bar \tau)\ ,
\end{equation}
where $k$ characterizes the various sectors ($k=0$ is the untwisted sector) and $l$ implements the orbifold projection in each sector\footnote{If we denote the orbifold group element by $g$, with $g^4=1$, then the projection operator takes the form $P=\frac{1}{4}(1+g+g^2+g^3)$.}. Furthermore, for our model the partition function will factorize into the following pieces (we omit writing the $\tau$ dependence for simplicity of the notation)
\begin{equation}
    {Z}[k,l]= {Z}_{\mathbb{R}^{1,4}}\,  {Z}_{S^1}[k,l]  {Z}_{T^4}[k,l]  {Z}_F[k,l]\,.
    \label{partition function first}
\end{equation}
Here ${Z}_{\mathbb{R}^{1,4}}$ is the contribution to the partition function from the non-compact bosons, ${Z}_{S^1}[k,l]$ and ${Z}_{T^4}[k,l]$ refer to the compact bosons on $S^1$ and $T^4$ respectively and ${Z}_F[k,l]$ is the fermionic contribution to the partition function.

Let us now discuss the above terms in detail. First of all, the contribution to the partition function from the non-compact bosons is  (we work in lightcone gauge)
\begin{equation}
    {Z}_{\mathbb{R}^{1,4}}= \left(\sqrt{\tau_2}\,\eta\,\bar{\eta}\right)^{-3}\,.
\end{equation}
This term is invariant under both $\mathcal{T}$ and $\mathcal{S}$ modular transformations.

The contribution to the partition function from the compact boson on $S^1$ is
\begin{equation}
     {Z}_{S^1}[k,l]=\frac{\mathcal{R}}{\sqrt{\alpha'}\sqrt{\tau_2}\,\eta\,\bar{\eta}}\sum_{n,w \in \mathbb{Z}}e^{-\frac{\pi \mathcal{R}^2}{\alpha'\tau_2}\left|n-\frac{l}{4}+\left(w+\frac{k}{4}\right)\tau\right|^2}\,
\end{equation}
or equivalently
\begin{equation}
{Z}_{S^1}[k,l]=  \frac{1}{\eta\,\bar{\eta}}\sum_{n,w \in \mathbb{Z}}e^{\frac{2\pi i n}{4}l}\, q^{\frac{\alpha'}{4}P_{R}^2(k)}\, \bar{q}^{\frac{\alpha'}{4}P_{L}^2(k)}\,,
\end{equation}
where
\begin{equation}
     P_{{L}/{R}}(k)=\frac{n}{\mathcal{R}}\pm  \frac{\big(w+\frac{k}{4}\big)\mathcal{R}}{\alpha'}\,.
\end{equation}
Note that the circle partition function consists of four building blocks: ${Z}_{S^1}[0,0]$, ${Z}_{S^1}[0,l], {Z}_{S^1}[k,0]$ and ${Z}_{S^1}[k,l]$. Of course, ${Z}_{S^1}[0,0]$ corresponds simply to a circle compactification and is invariant under both $\mathcal{T}$ and $\mathcal{S}$ modular transformations. The remaining blocks obey the following modular transformations
\begin{equation}
\begin{aligned}
   {Z}_{S^1}[k,l]\xrightarrow{\mathcal{T}}  \frac{\mathcal{R}}{\sqrt{\alpha'}\sqrt{\tau_2}\,\eta\,\bar{\eta}}&\sum_{n,w \in \mathbb{Z}}e^{-\frac{\pi \mathcal{R}^2}{\alpha'\tau_2}\left|(n+w)-\frac{l-k}{4}+\left(w+\frac{k}{4}\right)\tau\right|^2}= {Z}_{S^1}[k,l-k]\,,\\
{Z}_{S^1}[k,l]\xrightarrow{\mathcal{S}} \frac{\mathcal{R}}{\sqrt{\alpha'}\sqrt{\tau_2}\,\eta\,\bar{\eta}}&\sum_{n,w \in \mathbb{Z}}e^{-\frac{\pi \mathcal{R}^2}{\alpha'\tau_2}\left|w+\frac{k}{4}+\left(n+\frac{l}{4}\right)\tau\right|^2}={Z}_{S^1}[l,-k]\,.
\end{aligned}
\label{circle modular transf}
\end{equation}
Also, it is straightforward to verify the following properties
\begin{equation}
    {Z}_{S^1}[k,l]={Z}_{S^1}[-k,-l] ={Z}_{S^1}[-k+4,-l]={Z}_{S^1}[-k,-l+4]={Z}_{S^1}[-k+4,-l+4]\,.
    \label{circle ids}
\end{equation}
The above transformation rules \eqref{circle modular transf} and properties \eqref{circle ids} will be combined with similar ones from $Z_{T^4}[k,l]$ and ${Z}_F[k,l]$ to ensure modular invariance of the full partition function.

Finally, in order to compute the contribution to the partition function from the compact bosons on $T^4$ and from the fermions, which we denote by $\left({Z}_{T^4}{Z}_F\right)[k,l]$, we start from the untwisted sector and by performing modular $\mathcal{S}$ and $\mathcal{T}$ transformation we obtain pieces of the partition function in the twisted sectors. Similarly with the circle partition function, $\left({Z}_{T^4}{Z}_F\right)[0,0]$ is invariant under both $\mathcal{T}$ and $\mathcal{S}$ modular transformations. Note that $\left({Z}_{T^4}{Z}_F\right)[0,0]$ contains a Narain lattice $\Gamma^{4,4}(D_4)$.

In the untwisted sector, $\left({Z}_{T^4}{Z}_F\right)[0,l]$ can be computed by the general formula (see e.g. \cite{Gkountoumis:2023fym})
\begin{equation}
    \begin{aligned}
     \left({Z}_{T^4}{Z}_F\right)[0,l]=&\prod^4_{i=3}2 \sin(\pi l\tilde{u}_i)\, \frac{\bar{\eta}} {\bar{\vartheta}\Big[\psymbol{ \frac{1}{2}}{ -\frac{1}{2}+l\tilde{u}_i}\Big]} \frac{1}{2\bar{\eta}^4}\Big\{(\bar{\vartheta_3})^2\prod^4_{i=3}{{\bar{\vartheta}[\psymbol{0}{ -l\tilde{u}_i}]}}-(\bar{\vartheta_4})^2\prod^4_{i=3}\bar{\vartheta}[\psymbol{0}{ -\frac{1}{2}-l\tilde{u}_i}]\\
     &- (\bar{\vartheta_2})^2\prod^4_{i=3}{{\bar{\vartheta}[\psymbol{\frac{1}{2}}{ -l\tilde{u}_i}]}}-(\bar{\vartheta_1})^2\prod^4_{i=3}\bar{\vartheta}[\psymbol{\frac{1}{2}}{ -\frac{1}{2}-l\tilde{u}_i}]\Big\} \frac{1}{{\eta}^4}\Theta_{D_4}(\tau){\mathcal{Z}}_F[0,0]\,,
    \end{aligned}
\end{equation}
where ${\mathcal{Z}}_F[0,0]$ is the right-moving fermionic partition function, which reads
\begin{equation}
    {\mathcal{Z}}_F[0,0]=\frac{1}{2{\eta}^4}
      [(\vartheta_3)^4-(\vartheta_4)^4-(\vartheta_2)^4-(\vartheta_1)^4]\,.
\end{equation}
Recall that for the $\mathbb{Z}_4$ orbifold discussed in section \ref{the rank 4 model},  $\tilde{u}=(0,0,\tfrac{1}{4},\tfrac{1}{4})$. Let us now start with $\left({Z}_{T^4}{Z}_F\right)[0,1]$. We have
\begin{equation*}
\begin{aligned}
    \left({Z}_{T^4}{Z}_F\right)[0,1]=  \frac{\bar{\eta}^2}{ \bar{\vartheta}[\psymbol{{1}/{2}}{-{1}/{4}}]^2}\frac{1}{\bar{\eta}^4} &\left[(\bar{\vartheta_3})^2\bar{\vartheta}[\psymbol{0}{-{1}/{4}}]^2 - (\bar{\vartheta_4})^2\bar{\vartheta}[\psymbol{0}{-{3}/{4}}]^2 - (\bar{\vartheta_2})^2\bar{\vartheta}[\psymbol{{1}/{2}}{-{1}/{4}}]^2 -(\bar{\vartheta_1})^2\bar{\vartheta}[\psymbol{{1}/{2}}{-{3}/{4}}]^2\right]\\
  & \times\,\frac{1}{{\eta}^4}\Theta_{D_4}(\tau){\mathcal{Z}}_F[0,0] \quad \xrightarrow{\mathcal{S}}
  \end{aligned}
\end{equation*}
\begin{equation*}
\begin{aligned}
    \left({Z}_{T^4}{Z}_F\right)[1,0]=  \frac{i\bar{\eta}^2}{ \bar{\vartheta}[\psymbol{{1}/{4}}{-{1}/{2}}]^2}\frac{1}{\bar{\eta}^4} &\left[(\bar{\vartheta_3})^2\bar{\vartheta}[\psymbol{1/4}{0}]^2 +i(\bar{\vartheta_4})^2\bar{\vartheta}[\psymbol{1/4}{-{1}/{2}}]^2 - (\bar{\vartheta_2})^2\bar{\vartheta}[\psymbol{{3}/{4}}{0}]^2 +i(\bar{\vartheta_1})^2\bar{\vartheta}[\psymbol{{3}/{4}}{-{1}/{2}}]^2\right]\\
  & \times\,\frac{1}{2{\eta}^4}\Theta_{D_4^*}(\tau){\mathcal{Z}}_F[0,0] \quad \xrightarrow{\mathcal{T}}
  \end{aligned}
\end{equation*}
\begin{equation*}
\begin{aligned}
    \left({Z}_{T^4}{Z}_F\right)[1,3]=  \frac{i\bar{\eta}^2}{ \bar{\vartheta}[\psymbol{{1}/{4}}{{1}/{4}}]^2}\frac{1}{\bar{\eta}^4} &\left[(\bar{\vartheta_3})^2\bar{\vartheta}[\psymbol{1/4}{-3/4}]^2 -i(\bar{\vartheta_4})^2\bar{\vartheta}[\psymbol{1/4}{-{1}/{4}}]^2 - (\bar{\vartheta_2})^2\bar{\vartheta}[\psymbol{{3}/{4}}{-3/4}]^2 -i(\bar{\vartheta_1})^2\bar{\vartheta}[\psymbol{{3}/{4}}{-{1}/{4}}]^2\right]\\
  & \times\,\frac{1}{2{\eta}^4}\Theta_{D_4^*}(\tau+1){\mathcal{Z}}_F[0,0] \quad \xrightarrow{\mathcal{T}}
  \end{aligned}
\end{equation*}   
\begin{equation*}
\begin{aligned}
    \left({Z}_{T^4}{Z}_F\right)[1,2]=  \frac{i\bar{\eta}^2}{ \bar{\vartheta}[\psymbol{{1}/{4}}{0}]^2}\frac{1}{\bar{\eta}^4} &\left[(\bar{\vartheta_3})^2\bar{\vartheta}[\psymbol{1/4}{-1/2}]^2 -i(\bar{\vartheta_4})^2\bar{\vartheta}[\psymbol{1/4}{0}]^2 - (\bar{\vartheta_2})^2\bar{\vartheta}[\psymbol{{3}/{4}}{-1/2}]^2 -i(\bar{\vartheta_1})^2\bar{\vartheta}[\psymbol{{3}/{4}}{0}]^2\right]\\
  & \times\,\frac{1}{2{\eta}^4}\Theta_{D_4^*}(\tau){\mathcal{Z}}_F[0,0] \quad \xrightarrow{\mathcal{T}}
  \end{aligned}
\end{equation*}   
\begin{equation}
\begin{aligned}
     \left({Z}_{T^4}{Z}_F\right)[1,1]=  \frac{i\bar{\eta}^2}{ \bar{\vartheta}[\psymbol{{1}/{4}}{-1/4}]^2}\frac{1}{\bar{\eta}^4} &\left[(\bar{\vartheta_3})^2\bar{\vartheta}[\psymbol{1/4}{-1/4}]^2 +i(\bar{\vartheta_4})^2\bar{\vartheta}[\psymbol{1/4}{-3/4}]^2 - (\bar{\vartheta_2})^2\bar{\vartheta}[\psymbol{{3}/{4}}{-1/4}]^2 +i(\bar{\vartheta_1})^2\bar{\vartheta}[\psymbol{{3}/{4}}{-3/4}]^2\right]\\
  & \times\,\frac{1}{2{\eta}^4}\Theta_{D_4^*}(\tau+1){\mathcal{Z}}_F[0,0] \quad \xrightarrow{\mathcal{T}} \quad \left({Z}_{T^4}{Z}_F\right)[1,0]\,, 
  \end{aligned}
\end{equation} 
 where we also used that $\Theta_{D_4^*}(\tau+2)=\Theta_{D_4^*}(\tau)$. Now, by performing a modular $\mathcal{S}$ transformation on $\left({Z}_{T^4}{Z}_F\right)[1,2]$ we obtain
\begin{equation}
    \begin{aligned}
        \left({Z}_{T^4}{Z}_F\right)[2,3]=  \frac{-\bar{\eta}^2}{ \bar{\vartheta}[\psymbol{0}{1/4}]^2}\frac{1}{\bar{\eta}^4} &\left[(\bar{\vartheta_3})^2\bar{\vartheta}[\psymbol{1/2}{-3/4}]^2 +(\bar{\vartheta_4})^2\bar{\vartheta}[\psymbol{1/2}{-1/4}]^2 - (\bar{\vartheta_2})^2\bar{\vartheta}[\psymbol{0}{-3/4}]^2 +(\bar{\vartheta_1})^2\bar{\vartheta}[\psymbol{0}{-1/4}]^2\right]\\
  & \times\,\frac{1}{{\eta}^4}\Theta_{D_4}(\tau){\mathcal{Z}}_F[0,0]\,.
    \end{aligned}
    \label{z23 partition}
\end{equation}
In order to get some other pieces of the partition function effortless, we note that
\begin{equation}
    \left({Z}_{T^4}{Z}_F\right)[0,1]=\left({Z}_{T^4}{Z}_F\right)[0,3]\implies \left({Z}_{T^4}{Z}_F\right)[1,0]=\left({Z}_{T^4}{Z}_F\right)[3,0] \,.
    \label{equalities1}
\end{equation}
Then, it is also implied that
\begin{equation}
\begin{aligned}
    &\left({Z}_{T^4}{Z}_F\right)[1,3]=\left({Z}_{T^4}{Z}_F\right)[3,1]\,,\qquad\left({Z}_{T^4}{Z}_F\right)[1,2]=\left({Z}_{T^4}{Z}_F\right)[3,2]\,,\\
    &\left({Z}_{T^4}{Z}_F\right)[1,1]=\left({Z}_{T^4}{Z}_F\right)[3,3]\,,\qquad\left({Z}_{T^4}{Z}_F\right)[2,3]=\left({Z}_{T^4}{Z}_F\right)[2,1]\,.
    \end{aligned}
    \label{equalities2}
\end{equation}
Note that \eqref{equalities1} and \eqref{equalities2} are in agreement with the properties of the circle partition function \eqref{circle ids}. Also, from the above equations and \eqref{circle ids} we can see that the $k=1$ and $k=3$ sectors and equivalent. We will use all these later in order to check modular invariance easier.

Finally, in order to obtain the remaining pieces of the partition function we consider $\left({Z}_{T^4}{Z}_F\right)[0,2]$.
\begin{equation*}
      \left({Z}_{T^4}{Z}_F\right)[0,2]=2\left(\frac{\bar{\eta}}{\bar{\vartheta_2}}\right)^2\frac{1}{\bar{\eta}^4} [(\bar{\vartheta_3}\bar{\vartheta_4})^2-(\bar{\vartheta_4}\bar{\vartheta_3})^2-(\bar{\vartheta_2}\bar{\vartheta_1})^2-(\bar{\vartheta_1}\bar{\vartheta_2)^2}] 
       \frac{1}{{\eta}^4}\Theta_{D_4}(\tau){\mathcal{Z}}_F[0,0]\, \xrightarrow{\mathcal{S}}
\end{equation*}
\begin{equation*}
    \left({Z}_{T^4} {Z}_F\right)[2,0]= 2\left(\frac{\bar{\eta}}{\bar{\vartheta_4}}\right)^2\frac{1}{\bar{\eta}^4}  [(\bar{\vartheta_3}\bar{\vartheta_2})^2+(\bar{\vartheta_4}\bar{\vartheta_1})^2-(\bar{\vartheta_2}\bar{\vartheta_3})^2+(\bar{\vartheta_1}\bar{\vartheta_4)^2}] 
     \frac{1}{2{\eta}^4}\Theta_{D_4^*}(\tau){\mathcal{Z}}_F[0,0]\,\xrightarrow{\mathcal{T}}
\end{equation*}
\begin{equation}
    \left({Z}_{T^4}{Z}_F\right)[2,2]= 2\left(\frac{\bar{\eta}}{\bar{\vartheta_3}}\right)^2\frac{1}{\bar{\eta}^4}  [(\bar{\vartheta_3}\bar{\vartheta_1})^2+(\bar{\vartheta_4}\bar{\vartheta_2})^2-(\bar{\vartheta_2}\bar{\vartheta_4})^2+(\bar{\vartheta_1}\bar{\vartheta_3)^2}]  \frac{1}{2{\eta}^4}\Theta_{D_4^*}(\tau+1){\mathcal{Z}}_F[0,0]\,.
    \label{z22partition}
\end{equation}
Now we can put all pieces of the partition function together, and verify modular invariance for the model based on the $D_4$ lattice. Under a modular $\mathcal{T}$ transformation we find
\begin{equation}
    \begin{aligned}
   & \qquad \qquad Z[0,1]\,,Z[0,2]\,,Z[2,1]:\text{invariant}\,, \\
   & \qquad \qquad Z[1,0]\to Z[1,3]\to Z[1,2]\to Z[1,1]\to Z[1,0]\,,\\
   & \qquad \qquad Z[2,0]  \xleftrightarrow{} Z[2,2]\,.
    \end{aligned}
    \label{modular T orbit}
\end{equation}
Under a modular $\mathcal{S}$ transformation we find
\begin{equation}
    \begin{aligned}
   & Z[0,1]\xleftrightarrow{}Z[1,0]\,,\quad Z[0,2]\xleftrightarrow{}Z[2,0]\,,\\
   &Z[1,1]\xleftrightarrow{}Z[1,3]\,,\quad Z[1,2]\xleftrightarrow{}Z[2,1]\,,\\
   &Z[2,2]:\text{invariant}\,.
    \end{aligned}
    \label{modular S orbit}
\end{equation}
The transformations \eqref{modular T orbit} and \eqref{modular S orbit} together with \eqref{circle ids}, \eqref{equalities1} and \eqref{equalities2} ensure modular invariance.

\subsection{Spectrum}
\label{spectrumapp}
Here we present the spectrum of lightest states of the $\mathbb{Z}_4$ orbifolds discussed in section \ref{the rank 4 model}. This section is adjusted from \cite{Gkountoumis:2023fym}.

In order to obtain the spectrum of states we first rewrite the fermionic partition function in terms of infinite sums (cf. \eqref{14}); this is usually refereed to as bosonization. Then, since we are interested only in the lightest states, we expand the $\vartheta$-functions coming from the bosonic contributions as well as all the $\eta$-functions and we keep only the lowest order terms. Also, we omit writing down the $T^4$ lattice sums, as these are relevant for higher order terms. We start from the untwisted sector, in which we find (we omit irrelevant $\tau_2$ factors)
\begin{equation}
    {Z}[0,l]= (q\bar{q})^{-\frac{1}{2}}\sum_{n,w \in \mathbb{Z}}e^{\frac{\pi i n}{2}l}\, q^{\frac{\alpha'}{4}P_{R}^2(0)}\, (\bar{q})^{\frac{\alpha'}{4}P_{L}^2(0)}\sum_{{r},\tilde{{r}}} q^{\frac{1}{2}{r}^2}\,(\bar{q})^{\frac{1}{2}\tilde{r}^2}e^{\frac{\pi il}{2} (\tilde{r}_3+\tilde{r}_4)}\, \left(1+\cdots\right)\,.
    \label{N=6 massless untwisted}
\end{equation}
Here $\tilde{r}=(\tilde{r}_1,\tilde{r}_2,\tilde{r}_3,\tilde{r}_4)$ is an SO(8) weight vector with each component ${\tilde{r}}_i\in \mathbb{Z}$ in the NS-sector and ${\tilde{r}}\in \mathbb{Z}+\tfrac{1}{2}$ in the R-sector, and similarly for $r$. The GSO projection is $\sum_{i=1}^4 \tilde{r}_i \in 2\mathbb{Z}+1$ in the NS-sector and $\sum_{i=1}^4 \tilde{r}_i \in 2\mathbb{Z}$ in the R-sector (and similarly for $r$). The dots denote higher excited oscillator states. Also, in order to implement the orbifold projection we need to sum over $l$ and divide by the orbifold rank.

In \autoref{tablemasslessstates}, we present the NS and R-sector weight vectors for the states of the lowest level that survive the GSO projection (all of these are massless in the absence of momentum and/or winding modes). Furthermore, we table their representations under both the massless little group SO$(3)\approx\text{SU}(2)$ and the massive little group SO$(4)\approx\text{SU}(2)\times\text{SU}(2)$ in five dimensions. The latter is important when adding momenta or windings such that the states become massive.

\renewcommand{\arraystretch}{2}
\begin{table}[h!]
\centering
 \begin{tabular}{|c|c|c|c|}
    \hline
    Sector &  $\tilde{r}, {r}$  & SO(3) rep & SO(4) rep\\
    \hline
    \hline
   \multirow{3}{*}{NS}  & $(\underline{\pm 1,0},0,0)$ & $\textbf{3}\oplus \textbf{1}$ & $(\textbf{2},\textbf{2})$\\
    \cline{2-4}
    &$(0,0,\pm 1,0)$& 2$\,\times\,\textbf{1}$ & 2$\,\times\,(\textbf{1},\textbf{1})$\\
    \cline{2-4}
   & $(0,0,0,\pm 1)$ & 2$\,\times\,\textbf{1}$ & 2$\,\times\,(\textbf{1},\textbf{1})$\\
    \hline
    \hline
    \multirow{4}{*}{R}  & $(\pm\frac{1}{2},\pm\frac{1}{2},\frac{1}{2},\frac{1}{2})$ & $\textbf{2}$ & $(\textbf{2},\textbf{1})$\\
    \cline{2-4}
    & $(\pm\frac{1}{2},\pm\frac{1}{2},-\frac{1}{2},-\frac{1}{2})$ & $\textbf{2}$ & $(\textbf{2},\textbf{1})$\\
    \cline{2-4}
    & $(\underline{\frac{1}{2},-\frac{1}{2}},\frac{1}{2},-\frac{1}{2})$ & $\textbf{2}$ & $(\textbf{1},\textbf{2})$\\
    \cline{2-4}
    & $(\underline{\frac{1}{2},-\frac{1}{2}},-\frac{1}{2},\frac{1}{2})$ & $\textbf{2}$ & $(\textbf{1},\textbf{2})$\\
   \hline
    \end{tabular}
\captionsetup{width=.9\linewidth}
\caption{\textit{Here we write down all the weight vectors for states that are massless in the absence of momentum and/or winding modes, including their representations under the massless and massive little groups in 5D. We write down both left-moving and right-moving weight vectors, where underlining denotes permutations. This table is taken from} \cite{Gkountoumis:2023fym}.}
\label{tablemasslessstates}
\end{table}
\renewcommand{\arraystretch}{1}

We construct string states by tensoring the left and right-moving weight vectors from \autoref{tablemasslessstates}. We can see from \eqref{N=6 massless untwisted} that some states will carry a non-trivial orbifold charge, given by the phase $e^{\frac{\pi il}{2} (\tilde{r}_3+\tilde{r}_4)}$.
A charged state can survive the orbifold projection with the addition of appropriate momentum modes on the circle, such that the orbifold charge is cancelled. This also means that the state will become massive, since momentum modes contribute to the mass of a state. States with trivial charge survive the orbifold projection and remain massless.

For the construction of states, we use the rules
\begin{equation}
\begin{aligned}
\textbf{3}\otimes \textbf{3} = \textbf{5} \oplus \textbf{3} \oplus \textbf{1} \,, \qquad\quad \textbf{2}\otimes \textbf{2} = \textbf{3} \oplus \textbf{1} \,, \qquad\quad \textbf{3}\otimes \textbf{2} = \textbf{4} \oplus \textbf{2} \,,
\end{aligned}
\end{equation}
for tensoring SU$(2)$ representations.  In addition, we table the (massless and massive) representations that correspond to various supergravity fields in five dimensions in \autoref{table5Dfieldreps}.

\renewcommand{\arraystretch}{1.2}
\begin{table}[h!]
\centering
\begin{tabular}{|c|c|}
\hline
\;Massless field\; & \;SO$(3)$ rep\; \\ \hline\hline
$g_{\mu\nu}$ & \textbf{5} \\
$\psi_\mu$ & \textbf{4} \\
$A_\mu$ & \textbf{3} \\
$\chi$ & \textbf{2} \\
$\phi$ & \textbf{1} \\ \hline
\end{tabular}
\hspace{1.5cm}
\begin{tabular}{|c|c|}
\hline
\;\,Massive field\,\; & SO$(4)$ rep \\ \hline\hline
$B_{\mu\nu}^+$ / $B_{\mu\nu}^-$ & $(\textbf{3},\textbf{1})$ / $(\textbf{1},\textbf{3})$ \\
$\psi_\mu^+$ / $\psi_\mu^-$ & \;\:$(\textbf{2},\textbf{3})$ / $(\textbf{3},\textbf{2})$\:\; \\
$A_\mu$ & $(\textbf{2},\textbf{2})$ \\
$\chi^+$ / $\chi^-$ & $(\textbf{1},\textbf{2})$ / $(\textbf{2},\textbf{1})$ \\
$\phi$ & $(\textbf{1},\textbf{1})$ \\ \hline
\end{tabular}
\captionsetup{width=.83\linewidth}
\caption{\textit{Here we show the various massless and massive 5D fields and their representations under the appropriate little group. This table is taken from} \cite{Gkountoumis:2023fym}.}
\label{table5Dfieldreps}
\end{table}
\renewcommand{\arraystretch}{1}

Now, let us proceed with the construction of the massless spectrum in the untwisted sector. From \eqref{N=6 massless untwisted}, we observe that orbifold invariant states have to satisfy $\tilde{r}_3+\tilde{r}_4=0$ mod 4. By examining \autoref{tablemasslessstates} we find the following invariant states

NS-NS sector:
\begin{equation}
    \begin{aligned}
     (\underline{\pm 1,0},0,0) &\otimes (\underline{\pm 1,0},0,0)=\textbf{5}\oplus3\times\textbf{3}\oplus2\times\textbf{1}\\
     (\underline{\pm 1,0},0,0) &\otimes (0,0,\underline{\pm 1,0})=4\times\textbf{3}\oplus4\times\textbf{1}
    \end{aligned}
\end{equation}
NS-R sector:
\begin{equation}
    \begin{aligned}
        (\underline{\pm 1,0},0,0) &\otimes \pm (\pm\tfrac{1}{2},\pm\tfrac{1}{2},\tfrac{1}{2},\tfrac{1}{2})= 2\times \textbf{4}\oplus 4\times \textbf{2}  \\
        (\underline{\pm 1,0},0,0) &\otimes \pm(\underline{\tfrac{1}{2},-\tfrac{1}{2}},\tfrac{1}{2},-\tfrac{1}{2})=2\times \textbf{4}\oplus 4\times \textbf{2}
       \end{aligned}
\end{equation}
R-NS sector:
\begin{equation}
    \begin{aligned}
      \pm (\underline{\tfrac{1}{2},-\tfrac{1}{2}},\tfrac{1}{2},-\tfrac{1}{2})&\otimes (\underline{\pm 1,0},0,0) = 2\times \textbf{4}\oplus 4\times \textbf{2}\\
      \pm (\underline{\tfrac{1}{2},-\tfrac{1}{2}},\tfrac{1}{2},-\tfrac{1}{2}) &\otimes (0,0,\underline{\pm 1,0})=8\times \textbf{2}
    \end{aligned}
\end{equation}
R-R sector
\begin{equation}
    \begin{aligned}
      \pm (\underline{\tfrac{1}{2},-\tfrac{1}{2}},\tfrac{1}{2},-\tfrac{1}{2})&\otimes \pm (\pm\tfrac{1}{2},\pm\tfrac{1}{2},\tfrac{1}{2},\tfrac{1}{2})=4\times \textbf{3}\oplus 4 \times \textbf{1}\\
     \pm  (\underline{\tfrac{1}{2},-\tfrac{1}{2}},\tfrac{1}{2},-\tfrac{1}{2})&\otimes \pm (\underline{\tfrac{1}{2},-\tfrac{1}{2}},\tfrac{1}{2},-\tfrac{1}{2})=4 \times \textbf{3}\oplus 4\times \textbf{1}
     \end{aligned}
\end{equation}
All together, we find the graviton, 15 vectors, 14 scalars, 6 gravitini and 20 dilatini. These massless fields fit into the $\mathcal{N}=6$ gravity multiplet in $5D$. Regarding the notation, underlining denotes permutations, e.g. $(0,0,\underline{\pm 1,0})$ corresponds to $(0,0,1,0), (0,0,0,1), (0,0,-1,0)$ and $(0,0,0,-1)$. Now, we move on to the massive spectrum. For the construction of massive states, we take combinations from table \autoref{tablemasslessstates} that obtain a non-zero phase under the orbifold action and we cancel this phase by adding momentum modes on the circle. Whenever we add momentum and/or winding to a state, we denote it only on the left-movers by $(\tilde{r}_1,\tilde{r}_2,\tilde{r}_3,\tilde{r}_4;n,w)$. We find the following massive states

NS-NS sector:
\begin{equation}
    \begin{aligned}
     \pm(0,0,\underline{1,0};-1)&\otimes  (\underline{\pm 1,0},0,0)= 4\times(\textbf{2},\textbf{2})\\
    \pm (0,0,\underline{1,0};-1)&\otimes (0,0,\underline{\pm1,0})= 16\times(\textbf{1},\textbf{1})
    \end{aligned}
\end{equation}
NS-R sector:
\begin{equation}
    \begin{aligned}
     \pm (0,0,\underline{1,0};-1)&\otimes\pm(\pm\tfrac{1}{2},\pm\tfrac{1}{2},\tfrac{1}{2},\tfrac{1}{2})=8\times (\textbf{2},\textbf{1})\\
      \pm  (0,0,\underline{1,0};-1)&\otimes\pm(\underline{\tfrac{1}{2},-\tfrac{1}{2}},\tfrac{1}{2},-\tfrac{1}{2})=8\times (\textbf{1},\textbf{2})
    \end{aligned}
\end{equation}
R-NS sector:
\begin{equation}
    \begin{aligned}
    \pm   (\pm\tfrac{1}{2},\pm\tfrac{1}{2},\tfrac{1}{2},\tfrac{1}{2};-1)&\otimes(\underline{\pm 1,0},0,0)=2 \times (\textbf{3},\textbf{2})\oplus 2 \times (\textbf{1},\textbf{2})\\
      \pm (\pm\tfrac{1}{2},\pm\tfrac{1}{2},\tfrac{1}{2},\tfrac{1}{2};-1)&\otimes(0,0,\underline{\pm1,0})=8\times (\textbf{2},\textbf{1})
    \end{aligned}
\end{equation}
R-R sector:
\begin{equation}
    \begin{aligned}
     \pm (\pm\tfrac{1}{2},\pm\tfrac{1}{2},\tfrac{1}{2},\tfrac{1}{2};-1) &\otimes\pm(\pm\tfrac{1}{2},\pm\tfrac{1}{2},\tfrac{1}{2},\tfrac{1}{2})=4\times(\textbf{3},\textbf{1})\oplus 4 \times (\textbf{1},\textbf{1})\\
      \pm  (\pm\tfrac{1}{2},\pm\tfrac{1}{2},\tfrac{1}{2},\tfrac{1}{2};-1)&\otimes\pm (\underline{\tfrac{1}{2},-\tfrac{1}{2}},\tfrac{1}{2},-\tfrac{1}{2})=4\times(\textbf{2},\textbf{2})
    \end{aligned}
\end{equation}
In total, we find 2 gravitini (\textbf{3},\textbf{2}), 4 tensors (\textbf{3},\textbf{1}), 8 vectors (\textbf{2},\textbf{2}), 26 dilatini, $16\times(\textbf{2},\textbf{1})$ and $10\times(\textbf{1},\textbf{2})$, and 20 scalars (\textbf{1},\textbf{1}). All these fields have mass
\begin{equation}
    \frac{\alpha' m^2_{L}}{2}=\frac{\alpha' m^2_{R}}{2}=\frac{\alpha'}{4\mathcal{R}^2} \quad\Rightarrow \quad m= \left|\frac{1}{\mathcal{R}}\right|\,,
\end{equation}
due to the contribution of the $n=\pm1$ momentum mode on the circle, and fit into a complex (1,2) BPS supermultiplet in the representations
\begin{equation}
(\textbf{3},\textbf{2})\oplus2\times(\textbf{3},\textbf{1})\oplus4\times(\textbf{2},\textbf{2})\oplus5\times(\textbf{1},\textbf{2})\oplus8\times(\textbf{2},\textbf{1})\oplus10\times(\textbf{1},\textbf{1})\,.
\label{bpsmulti}
\end{equation}
We mention here that we can construct Kaluza-Klein towers along the $S^1$ by adding a trivial phase $e^{(\frac{\pi i l}{2})4\mathbb{Z}}$ to all states. Then, the orbifold untwisted spectrum matches exactly with the Scherk-Schwarz  supergravity one found in \cite{Hull:2020byc} (with the identification $\mathcal{R}=4R$, where $R$ is the radius of the Scherk-Schwarz circle).

We continue with the spectrum of lightest states in the twisted sectors, and we start from the $k=1$ sector. As before, we expand the partition function in powers of $q\bar{q}$ and focus on the lowest order terms. We have
\begin{equation}
     Z[1,l]=q^{-\frac{1}{2}}(\bar{q})^{-\frac{5}{16}}\,\sum_{{n,w\in \mathbb{Z}}}e^{\frac{\pi i n}{2}l} q^{\frac{\alpha'}{4}P_{R}^2(1)} (\bar{q})^{\frac{\alpha'}{4}P_{L}^2(1)}\,\sum_{{r},\tilde{r}}  q^{\frac{1}{2}{r}^2}(\bar{q})^{\frac{1}{2}(\tilde{r}+\tilde{u})^2} e^{\frac{\pi il}{2}(\tilde{r}_3+\tilde{r}_4+1)}(1+\cdots)\,.
\end{equation}
 The weight vectors for the lightest right-moving states are, again, given in \autoref{tablemasslessstates}. The weight vectors for the lightest left-moving states are listed in \autoref{k=1 N=6 states}. 
\renewcommand{\arraystretch}{2}
\begin{table}[h!]
\centering
 \begin{tabular}{|c|c|c|}
    \hline
    Sector &  $\tilde{r}$ & SO(4) rep\\
    \hline
    \hline
  NS & $(0,0,\underline{-1,0})$ & 2$\,\times\,(\textbf{1},\textbf{1})$\\
  \hline
   R  & $(\pm\frac{1}{2},\pm\frac{1}{2},-\frac{1}{2},-\frac{1}{2})$ & $(\textbf{2},\textbf{1})$\\
   \hline
    \end{tabular}
\captionsetup{width=.9\linewidth}
\caption{\textit{The weight vectors of the lightest left-moving states in the $k=1$ twisted sector, and their representations under the massive little group in 5D.}}
\label{k=1 N=6 states}
\end{table}
\renewcommand{\arraystretch}{1}

In the $k=1$ sector, orbifold invariant states have to satisfy $\tilde{r}_3+\tilde{r}_4+1=0$ mod 4. We list below the states that we find in each sector

NS-NS sector:
\begin{equation}
\begin{aligned}
 (0,0,\underline{-1,0}) &\otimes (\underline{\pm 1,0},0,0)= 2 \times (\textbf{2},\textbf{2})\\
     (0,0,\underline{-1,0}) &\otimes  (0,0,\underline{\pm1,0})= 8\times (\textbf{1},\textbf{1})
\end{aligned}
\end{equation}
NS-R sector:
\begin{equation}
    \begin{aligned}
       (0,0,\underline{-1,0})&\otimes\pm (\pm\tfrac{1}{2},\pm\tfrac{1}{2},\tfrac{1}{2},\tfrac{1}{2})=4\times (\textbf{2},\textbf{1})\\
         (0,0,\underline{-1,0})&\otimes\pm(\underline{\tfrac{1}{2},-\tfrac{1}{2}},\tfrac{1}{2},-\tfrac{1}{2})=4\times (\textbf{1},\textbf{2})
    \end{aligned}
\end{equation}
R-NS sector:
\begin{equation}
    \begin{aligned}
        (\pm\tfrac{1}{2},\pm\tfrac{1}{2},-\tfrac{1}{2},-\tfrac{1}{2})&\otimes(\underline{\pm 1,0},0,0)=(\textbf{3},\textbf{2})\oplus (\textbf{1},\textbf{2})\\
       (\pm\tfrac{1}{2},\pm\tfrac{1}{2},-\tfrac{1}{2},-\tfrac{1}{2})&\otimes(0,0,\underline{\pm1,0})=4\times (\textbf{2},\textbf{1})
    \end{aligned}
\end{equation}
R-R sector:
\begin{equation}
    \begin{aligned}
            (\pm\tfrac{1}{2},\pm\tfrac{1}{2},-\tfrac{1}{2},-\tfrac{1}{2}) &\otimes\pm (\pm\tfrac{1}{2},\pm\tfrac{1}{2},\tfrac{1}{2},\tfrac{1}{2})=2\times (\textbf{3},\textbf{1})\oplus 2 \times (\textbf{1},\textbf{1})\\
        (\pm\tfrac{1}{2},\pm\tfrac{1}{2},-\tfrac{1}{2},-\tfrac{1}{2})&\otimes \pm (\underline{\tfrac{1}{2},-\tfrac{1}{2}},\tfrac{1}{2},-\tfrac{1}{2})=2\times (\textbf{2},\textbf{2})
    \end{aligned}
\end{equation}
In total we find 1 gravitino (\textbf{3},\textbf{2}), 2 tensors (\textbf{3},\textbf{1}), 4 vectors (\textbf{2},\textbf{2}), 13 dilatini, $8\times(\textbf{2},\textbf{1})$ and $5\times(\textbf{1},\textbf{2})$, and 10 scalars (\textbf{1},\textbf{1}). All these fields have mass $\left|{\mathcal{R}}/{4\alpha'}\right|$ due to the $\tfrac{1}{4}$-winding on the circle. In the $k=3$ sector we find exactly the same spectrum (in the $k=3$ sector the lightest states are those with winding number $w=-1$). Together, the fields from both the $k=1$ and $k=3$ sectors fit into a complex (1,2) BPS supermultiplet, as in \eqref{bpsmulti}.

Finally, in the $k=2$ sector we have
\begin{equation}
     Z[2,l]=2  q^{-\frac{1}{2}}(\bar{q})^{-\frac{1}{4}}\,\sum_{{n,w\in \mathbb{Z}}}e^{\frac{\pi i n}{2}l} q^{\frac{\alpha'}{4}P_{R}^2(2)} (\bar{q})^{\frac{\alpha'}{4}P_{L}^2(2)}\,\sum_{{r},\tilde{r}}  q^{\frac{1}{2}{r}^2}(\bar{q})^{\frac{1}{2}(\tilde{r}+2\tilde{u})^2} e^{\frac{\pi i l}{2}(\tilde{r}_3+\tilde{r}_4-1)}(1+\cdots)\,,
\end{equation}
where the overall factor of $2$, can be read of from the relevant pieces of the partition function \eqref{z23 partition} and \eqref{z22partition}, after bosonization (also, recall that $Z[2,1]=Z[2,3]$). Regarding the spectrum of the $k=2$ sector, the weight vectors of the lightest states coincide with those in the $k=1$ sector. However, in the $k=2$ sector states with $\tilde{r}=(0,0,\underline{-1,0})$ or $\tilde{r}=(\pm\frac{1}{2},\pm\frac{1}{2},-\frac{1}{2},-\frac{1}{2})$  survive the orbifold projection only if suitable combination of momentum modes, winding modes and $T^4$ right-moving momenta are added to these states. These are higher excited stringy states, which we omit writing down.

\subsection{Conventions and identities}
\label{modular functions}
The Dedekind $\eta$-function is defined as
\begin{equation}
   \eta(\tau)\equiv q^{\frac{1}{24}} \prod_{n=1}^{\infty}(1-q^n)\,,\hspace{0.5cm}q=e^{2\pi i \tau}\,.
\end{equation}
The Jacobi $\vartheta$-function with characteristics $\alpha,\beta$ is given by
\begin{equation}
     \vartheta[\psymbol{ \alpha}{ \beta}](\tau)=\sum_{n\in \mathbb{Z}}q^{\frac{1}{2}(n+\alpha)^2}e^{2\pi i(n+\alpha)\beta}\,.
     \label{14}
\end{equation}
For $-\tfrac{1}{2}\leq \alpha, \beta \leq \tfrac{1}{2}$ there is also a product representation of the $\vartheta$-function, which reads 
\begin{equation}
    \vartheta[\psymbol{ \alpha}{ \beta}](\tau)=\eta(\tau)e^{2\pi i\alpha\beta}q^{\frac{1}{2}\alpha^2-\frac{1}{24}}\prod_{n=1}^{\infty}(1+q^{n+\alpha-\frac{1}{2}}e^{2\pi i\beta})(1+q^{n-\alpha-\frac{1}{2}}e^{-2\pi i\beta})\,.
    \label{product representation}
\end{equation}
Particular $\vartheta$-functions that appear often are 
\begin{equation}
    \vartheta[\psymbol{ 0}{ 0}](\tau)\equiv\vartheta_3(\tau)\,,\hspace{0,2cm} \vartheta[\psymbol{ 0}{ \frac{1}{2}}](\tau)\equiv\vartheta_4(\tau)\,,\hspace{0,2cm} \vartheta[\psymbol{ \frac{1}{2}}{ 0}](\tau)\equiv\vartheta_2(\tau)\,,\hspace{0,2cm} \vartheta[\psymbol{ \frac{1}{2}}{ \frac{1}{2}}](\tau)\equiv\vartheta_1(\tau)\,.
\end{equation}
In addition, two useful $\vartheta$-function identities are
\begin{equation}
    \vartheta[\psymbol{ -\alpha}{ -\beta}](\tau)=  \vartheta[\psymbol{ \alpha}{ \beta}](\tau)\,,\hspace{0,3 cm}  \vartheta[\psymbol{ \alpha+m}{ \beta+n}](\tau)= e^{2\pi i n \alpha}\,\vartheta[\psymbol{ \alpha}{ \beta}](\tau)\,,\,\,m,n \in \mathbb{Z}\,.
\end{equation}
The modular transformations are defined as: $\mathcal{T}\equiv \tau\to \tau+1$ and $\mathcal{S}\equiv \tau \to -{1}/{\tau}$. The Dedekind $\eta$-function transforms as follows 
\begin{equation}
\begin{aligned}
    &\eta(\tau+1)= e^{{\pi i}/{12}}\,\eta(\tau)\,,\\
   & \eta(-{1}/{\tau}) = \sqrt{-i\tau}\,\eta(\tau)\,.
    \end{aligned}
\end{equation}
Note that under both $\mathcal{T}$ and $\mathcal{S}$ transformations the combination $\sqrt{\tau_2}\,\eta(\tau)\,\bar{\eta}(\bar{\tau})$ is invariant. Under modular transformations, the Jacobi $\theta$-function transforms as follows
\begin{equation}
    \begin{aligned}
      & \vartheta[\psymbol{ \alpha}{ \beta}] (\tau+1)=e^{-\pi i(\alpha^2-\alpha)} \, \vartheta[\psymbol{ \alpha}{\alpha+\beta-\frac{1}{2}}](\tau)\,,\\
       & \vartheta[\psymbol{ \alpha}{ \beta}] (-1/\tau) = \sqrt{-i\tau}\,  e^{2\pi i \alpha \beta}\, \vartheta[\psymbol{ -\beta}{ \alpha}](\tau)\,.
    \end{aligned}
\end{equation}
The theta series of the SO(2n) root lattice $D_n$ is
\begin{equation}
    \Theta_{D_n}(\tau)= \frac{1}{2}\left(\vartheta_3(\tau)^n+\vartheta_4(\tau)^n\right)\,.
\end{equation}
Note the isomorphism $D_2\cong A_1\oplus A_1$. Finally, for a $d$-dimensional lattice $\Lambda$ and its dual $\Lambda^*$ the following expression holds
\begin{equation}
    \Theta_{\Lambda}({-1}/{\tau})= \frac{(-i\tau)^{\frac{d}{2}}}{\text{Vol}(\Lambda)} \Theta_{\Lambda^*}({\tau})\,.
\end{equation}

\bibliographystyle{unsrt}
\bibliography{bib}

\begin{thebibliography}{10}

\bibitem{Gkountoumis:2023fym}
George Gkountoumis, Chris Hull, Koen Stemerdink, and Stefan Vandoren.
\newblock {Freely acting orbifolds of type IIB string theory on T$^{5}$}.
\newblock {\em JHEP}, 08:089, 2023.

\bibitem{Rohm:1983aq}
Ryan Rohm.
\newblock {Spontaneous Supersymmetry Breaking in Supersymmetric String Theories}.
\newblock {\em Nucl. Phys. B}, 237:553--572, 1984.

\bibitem{Kounnas:1988ye}
Costas Kounnas and Massimo Porrati.
\newblock {Spontaneous Supersymmetry Breaking in String Theory}.
\newblock {\em Nucl. Phys. B}, 310:355--370, 1988.

\bibitem{Ferrara:1987es}
Sergio Ferrara, Costas Kounnas, and Massimo Porrati.
\newblock {Superstring Solutions With Spontaneously Broken Four-dimensional Supersymmetry}.
\newblock {\em Nucl. Phys. B}, 304:500--512, 1988.

\bibitem{Ferrara:1988jx}
Sergio Ferrara, Costas Kounnas, Massimo Porrati, and Fabio Zwirner.
\newblock {Superstrings with Spontaneously Broken Supersymmetry and their Effective Theories}.
\newblock {\em Nucl. Phys. B}, 318:75--105, 1989.

\bibitem{Kiritsis:1997ca}
Elias Kiritsis and Costas Kounnas.
\newblock {Perturbative and nonperturbative partial supersymmetry breaking: N=4 ---\ensuremath{>} N=2 ---\ensuremath{>} N=1}.
\newblock {\em Nucl. Phys. B}, 503:117--156, 1997.

\bibitem{bianchi2022perturbative}
Massimo Bianchi, Guillaume Bossard, and Dario Consoli.
\newblock {Perturbative higher-derivative terms in N= 6 asymmetric orbifolds}.
\newblock {\em Journal of High Energy Physics}, 2022(6):1--96, 2022.

\bibitem{Ferrara:1989nm}
Sergio Ferrara and Costas Kounnas.
\newblock {Extended Supersymmetry in Four-dimensional Type {II} Strings}.
\newblock {\em Nucl. Phys. B}, 328:406--438, 1989.

\bibitem{Bianchi:2008cj}
Massimo Bianchi.
\newblock {Bound-states of D-branes in L-R asymmetric superstring vacua}.
\newblock {\em Nucl. Phys. B}, 805:168--181, 2008.

\bibitem{Bianchi:2010aw}
Massimo Bianchi.
\newblock {On $\mathcal{R}^{4}$ terms and MHV amplitudes in $\mathcal{N}$ = 5,6 supergravity vacua of Type II superstrings}.
\newblock {\em Adv. High Energy Phys.}, 2011:479038, 2011.

\bibitem{Narain:1986qm}
Kumar Narain, M.Hossein Sarmadi, and Cumrun Vafa.
\newblock {Asymmetric Orbifolds}.
\newblock {\em Nucl. Phys.}, B288:551, 1987.

\bibitem{narain1991asymmetric}
Kumar Narain, M.Hossein Sarmadi, and Cumrun Vafa.
\newblock Asymmetric orbifolds: Path integral and operator formulations.
\newblock {\em Nuclear Physics B}, 356(1):163--207, 1991.

\bibitem{Gkountoumis:2024dwc}
George Gkountoumis, Chris Hull, and Stefan Vandoren.
\newblock {Exact moduli spaces for $\mathcal{N}=2$, $D=5$ freely acting orbifolds}.
\newblock {\em arXiv preprint hep-th/2403.05650}, 2024.

\bibitem{narain1989new}
Kumar Narain.
\newblock New heterotic string theories in uncompactified dimensions $<$ 10.
\newblock In {\em Current Physics--Sources and Comments}, volume~4, pages 246--251. Elsevier, 1989.

\bibitem{lerchie1989lattices}
Wolfgang Lerchie, Adrianus Schellekens, and Nicholas Warner.
\newblock Lattices and strings.
\newblock {\em Physics Reports}, 177(1-2):1--140, 1989.

\bibitem{vafa1986modular}
Cumrun Vafa.
\newblock Modular invariance and discrete torsion on orbifolds.
\newblock {\em Nuclear Physics B}, 273(3-4):592--606, 1986.

\bibitem{Baykara:2023plc}
Zihni~Kaan Baykara, Yuta Hamada, Houri-Christina Tarazi, and Cumrun Vafa.
\newblock {On the String Landscape Without Hypermultiplets}.
\newblock 9 2023.

\bibitem{Harvey:2017rko}
Jeffrey Harvey and Gregory Moore.
\newblock {An Uplifting Discussion of T-Duality}.
\newblock {\em JHEP}, 05:145, 2018.

\bibitem{Cremmer:1980gs}
Eugene Cremmer.
\newblock {Supergravities in 5 Dimensions}.
\newblock In {\em Supergravities in diverse dimensions. Volume 1}, 1989.

\bibitem{Scherk:1978ta}
Joel Scherk and John Schwarz.
\newblock {Spontaneous Breaking of Supersymmetry Through Dimensional Reduction}.
\newblock {\em Phys. Lett.}, 82B:60--64, 1979.

\bibitem{Scherk:1979zr}
Joel Scherk and John Schwarz.
\newblock {How to Get Masses from Extra Dimensions}.
\newblock {\em Nucl. Phys.}, B153:61--88, 1979.

\bibitem{Cremmer:1979uq}
Eugene Cremmer, Joel Scherk, and John Schwarz.
\newblock {Spontaneously Broken N=8 Supergravity}.
\newblock {\em Phys. Lett.}, 84B:83--86, 1979.

\bibitem{Hull:2020byc}
Chris Hull, Eric Marcus, Koen Stemerdink, and Stefan Vandoren.
\newblock {Black holes in string theory with duality twists}.
\newblock {\em JHEP}, 07:086, 2020.

\bibitem{Awada:1985ep}
Moustafa Awada and Paul Townsend.
\newblock {$N=4$ Maxwell-Einstein supergravity in five dimensions and its SU(2) gauging}.
\newblock {\em Nucl. Phys. B}, 255:617--632, 1985.

\bibitem{de2001triples}
Jan de~Boer, Robbert Dijkgraaf, Kentaro Hori, Arjan Keurentjes, John Morgan, David Morrison, and Savdeep Sethi.
\newblock Triples, fluxes, and strings.
\newblock {\em arXiv preprint hep-th/0103170}, 2001.

\bibitem{bedroya2022compactness}
Alek Bedroya, Yuta Hamada, Miguel Montero, and Cumrun Vafa.
\newblock Compactness of brane moduli and the string lamppost principle in d $>$ 6.
\newblock {\em Journal of High Energy Physics}, 2022(2):1--31, 2022.

\bibitem{fraiman2023unifying}
Bernardo Fraiman and H{\'e}ctor Freitas.
\newblock {Unifying the 6D $\mathcal{N}$=(1, 1) String Landscape}.
\newblock {\em Journal of High Energy Physics}, 2023(2209.06214):1--42, 2023.

\bibitem{Dolivet:2007sz}
Yacine Dolivet, Bernard Julia, and Costas Kounnas.
\newblock {Magic N=2 supergravities from hyper-free superstrings}.
\newblock {\em JHEP}, 02:097, 2008.

\bibitem{Anastasopoulos:2009kj}
Pascal Anastasopoulos, Massimo Bianchi, Jose Morales, and Gianfranco Pradisi.
\newblock {(Unoriented) T-folds with few T's}.
\newblock {\em JHEP}, 06:032, 2009.

\bibitem{Condeescu:2013yma}
Cezar Condeescu, Ioannis Florakis, Costas Kounnas, and Dieter L\"ust.
\newblock {Gauged supergravities and non-geometric Q/R-fluxes from asymmetric orbifold CFT`s}.
\newblock {\em JHEP}, 10:057, 2013.

\bibitem{Hull:2017llx}
Chris Hull, Dan Israël, and Alessandra Sarti.
\newblock {Non-geometric Calabi-Yau Backgrounds and K3 automorphisms}.
\newblock {\em JHEP}, 11:084, 2017.

\bibitem{Gautier:2019qiq}
Yoan Gautier, Chris Hull, and Dan Isra\"el.
\newblock {Heterotic/type II Duality and Non-Geometric Compactifications}.
\newblock {\em JHEP}, 10:214, 2019.

\bibitem{Israel:2013wwa}
Dan Isra\"el and Vincent Thi\'ery.
\newblock {Asymmetric Gepner models in type II}.
\newblock {\em JHEP}, 02:011, 2014.

\bibitem{Gunaydin:1983rk}
Murat Gunaydin, German Sierra, and Paul Townsend.
\newblock {Exceptional Supergravity Theories and the MAGIC Square}.
\newblock {\em Phys. Lett. B}, 133:72--76, 1983.

\bibitem{Antoniadis:2003sw}
Ignatios Antoniadis, Ruben Minasian, Stefan Theisen, and Pierre Vanhove.
\newblock {String loop corrections to the universal hypermultiplet}.
\newblock {\em Class. Quant. Grav.}, 20:5079--5102, 2003.

\bibitem{Robles-Llana:2006vct}
Daniel Robles-Llana, Frank Saueressig, and Stefan Vandoren.
\newblock {String loop corrected hypermultiplet moduli spaces}.
\newblock {\em JHEP}, 03:081, 2006.

\bibitem{Robles-Llana:2006hby}
Daniel Robles-Llana, Martin Rocek, Frank Saueressig, Ulrich Theis, and Stefan Vandoren.
\newblock {Nonperturbative corrections to 4D string theory effective actions from SL(2,Z) duality and supersymmetry}.
\newblock {\em Phys. Rev. Lett.}, 98:211602, 2007.

\bibitem{Robles-Llana:2007bbv}
Daniel Robles-Llana, Frank Saueressig, Ulrich Theis, and Stefan Vandoren.
\newblock {Membrane instantons from mirror symmetry}.
\newblock {\em Commun. Num. Theor. Phys.}, 1:681--711, 2007.

\bibitem{Alexandrov:2008gh}
Sergei Alexandrov, Boris Pioline, Frank Saueressig, and Stefan Vandoren.
\newblock {D-instantons and twistors}.
\newblock {\em JHEP}, 03:044, 2009.

\bibitem{Alexandrov:2021shf}
Sergei Alexandrov, Ashoke Sen, and Bogdan Stefa\'nski.
\newblock {D-instantons in Type IIA string theory on Calabi-Yau threefolds}.
\newblock {\em JHEP}, 11:018, 2021.

\bibitem{Alexandrov:2021dyl}
Sergei Alexandrov, Ashoke Sen, and Bogdan Stefa\'nski.
\newblock {Euclidean D-branes in type IIB string theory on Calabi-Yau threefolds}.
\newblock {\em JHEP}, 12:044, 2021.

\bibitem{Alexandrov:2023hiv}
Sergei Alexandrov and Khalil Bendriss.
\newblock {Hypermultiplet metric and NS5-instantons}.
\newblock 9 2023.

\end{thebibliography}

\end{document}